\documentstyle[aps,eqsecnum]{revtex}
\begin{document}
\twocolumn[\hsize\textwidth\columnwidth\hsize\csname
@twocolumnfalse\endcsname
\title{Analytic derivation of the map of null rays passing near 
a naked singularity}
\author{Takahiro Tanaka${}^{1}$ 
 and T. P. Singh$^{1,2}$ \\ $~$}
\address{$^{1}$Yukawa Institute for Theoretical Physics, 
 Kyoto University, Kyoto 606-8502, Japan}
\address{$^{2}$Tata Institute of Fundamental Research, 
 Homi Bhabha Road, Mumbai 400 005, India.}

\maketitle

\thispagestyle{empty}
\begin{abstract}
Recently the energy emission from a naked singularity forming in
spherical dust collapse has been investigated.
This  radiation is due to the particle creation in a curved 
spacetime. In this discussion, the central role is played 
by the mapping formula between the incoming and the outgoing 
null coordinates. For the self-similar model, this mapping formula 
has been derived analytically. But for the model with 
$C^{\infty}$ density profile, the mapping formula has been 
obtained only numerically. 
In the present paper, we argue that the singular 
nature of the mapping is determined by the local geometry around 
the point at which the singularity is first formed. 
If this is the case, it would be natural to expect that 
the mapping formula can be derived analytically.  
In the present paper, 
we analytically rederive the same mapping formula 
for the model with $C^{\infty}$ density profile that 
has been earlier derived using a numerical technique. \\
{PACS 04.20.Dw, 04.70.Dy~~~~~~~YITP-00-57}
\end{abstract}

\vskip2pc]

\baselineskip10.8pt

\section{Introduction}
There are many known examples in the literature showing that naked
singularities can form from regular initial data, in the classical 
gravitational collapse of a bounded object. Quantum effects analogous
to Hawking radiation are expected to play an important role in
determining the final outcome of such a collapse. A study of such
effects has received some attention in recent years. The most notable
feature of these studies is that if a massless scalar field is
quantized on the background of a star forming a naked singularity, the
outgoing quantum flux of the scalar field can be calculated in the
geometric optics approximation\cite{fop,SiVaWi,Har1,Har2,SiVa}, 
and in 2-d models\cite{His,bsv,bsv2}.
This quantum flux can be shown to diverge on the Cauchy horizon.
An interpretation of this divergence has recently been given 
by \cite{HINSTV}.

The calculation of the quantum flux in the geometric optics
approximation crucially relies on the determination of the map between
ingoing null rays coming in from ${\cal I}^{-}$ and outgoing null rays
going to ${\cal I}^{+}$. This map can be calculated in a
straightforward manner for the model of self-similar spherical dust
collapse. However, when the self-similarity condition is dropped, the
calculation becomes difficult and the map has so far only been
determined numerically.

In this paper we argue that this map is determined, not by the global
geometry, but by the local geometry around the point at which the
singularity is first formed. It is then possible to determine the map 
analytically.

In this paper, we solve the radial null rays in 
the Tolman-Bondi spacetime approximately. These null rays naturally 
defines a map between the incoming and outgoing null 
coordinates, $u$ and $v$. As noted above, this map plays a crucial
role in calculating 
the energy emission due to the quantum effect in curved 
spacetime in the geometrical optics approximation. 

For the self-similar model with a stellar surface on which 
the background solution is matched to the Schwarzschild metric, 
this map was determined analytically\cite{SiVaWi}.  
It was shown that the map takes the form of a power law : 
$u_0-u\propto (v_0-v)^\gamma$, where $u=u_0$ and $v=v_0$ are 
outgoing and incoming null rays passing through the point 
at which the singularity is first formed. 

On the other hand, such a map was not derived analytically 
for the more generic so-called $\rho_2$-models for which we have 
non-vanishing second derivative of the density profile at the center 
before the formation of naked singularity. 
Numerically, the map was shown\cite{Har1,Har2} to be given as 
$d^2(u_0-u)/dv^2\propto (v_0-v)^{-1/2}$. 

In this paper, we would like to show two things. 
First we shall point out that it is not necessary to study the 
global solution for radial null geodesics far from the singularity  
for the purpose of investigating the singular behavior of the map. 
We shall show that the basic feature of the map can be extracted 
just by considering a small region near 
the point at which the singularity is first formed.  

To demonstrate this, we propose an alternative map. 
Let us consider sending radial incoming null rays from an 
observer on a comoving shell. These null rays are reflected 
at the center, and come back to the same comoving observer. 
A radial null geodesic crosses a comoving 
shell located at a fixed comoving radius $r$ twice before and after
the reflection at the center. 
Thus these null rays define a map between the sending time and 
the receiving time measured by the proper time for the comoving 
observer. 
We shall show that this map possesses the same structure of singularity 
as the previously investigated map between null coordinates. 
In defining this map, we will see that the radius of the comoving 
shell can be chosen arbitrary small.  
One would expect such a result because there is no singular 
feature in the map between ticks on a comoving shell at a finite 
distance and that of the null coordinates naturally defined at infinity. 

Secondly, using the same map, we show that the structure of 
singularity for $\rho_2$-models can be obtained analytically. 

Throughout this paper, we use the units $8\pi G=1$ and $c=1$. 

\section{Self-Similar case}
The marginally bound Tolman-Bondi dust collapse is described by the
metric
\begin{equation}
ds^{2}=  dt^{2} -  R'^{2} dr^{2} - R^{2}  d\Omega^{2}
\end{equation}
and the evolution of 
the circumferential radius $R(t,r)$ and 
the density $\rho(r,t)$ is determined by the Einstein equations as 
\begin{equation}
\dot{R}^{2}={F(r) \over R}\, , 
\qquad \rho = {F'\over R^{2} R'}\, . 
\label{basic}
\end{equation}
In the above equations, a prime and a dot denote 
a derivative with respect to 
$r$ and $t$, respectively. 
Here $F(r)$ is the mass function which in case of the self-similar
model is given by $F(r)=\lambda r$ when the scaling is $R=r$ at the
singular epoch $t=0$. Collapse is assumed to begin at some epoch $t<0$.

This self-similar model was discussed in Ref.\cite{SiVaWi}.  
The evolution of the circumferential radius is given by 
\begin{equation}
R^{3/2}=r^{3/2}\left(1-{a t\over r}\right)\, , 
\end{equation}
where we have introduced $a={3\over 2}\sqrt{\lambda}$.
In this model, the initial density profile 
at $t=t_{in}<0$ near the center is of the
form
\begin{equation}
\rho = \rho_{0} + \rho_{3} R^{3} + \rho_{6} R^{6} +\cdots .
\end{equation}
Here the coefficient $\rho_{3}=-(16/3)a^{-3}(-t_{in})^5$ is 
negative, in which case the central singularity is known to be 
globally naked for a range of values of $\lambda$.
The values of higher order coefficients such as $\rho_{6}$ and $\rho_{9}$ 
are accordingly tuned so as to realize a self-similar solution. 

The equation which determines the 
incoming (upper sign) and outgoing (lower sign) radial null geodesics 
on this background geometry is given by 
\begin{equation}
 \mp {dt\over dr}=R'
    ={1-{1\over 3}{a t/r}
           \over (1-a t/r)^{1/3}}\,. 
\end{equation}
Following Ref.\cite{SiVaWi}, we introduce the variable 
\begin{equation}
y=\left(1-a t/r\right)^{1/3}. 
\end{equation}
Then the above equation is rewritten as 
\begin{equation}
{dr\over r\, dy}=-{9y^3 \over  3y^4\mp a y^3-3y\mp 2a} 
=-\sum_{i=1}^{4}{3A_i^{\pm}\over (y-\alpha_i^{\pm})}\, , 
\label{sseq}
\end{equation}
where $\alpha_i^{\pm}$ are the roots of $3y^4\mp a y^3-3y\mp 2a=0$, 
and the last equality defines the coefficients $A_i^{\pm}$,
which satisfy $\sum_{i=1}^{4}A_i^{\pm}=1$. 
Equation (\ref{sseq}) is integrated to obtain the solution,
\begin{equation}
 {r\over r_{0\pm}}=\prod_{i=1}^4 (y-\alpha_i^{\pm})^{-3 A_i^{\pm}}. 
\end{equation}
Here $r_{0\pm}$ is the integration constant which parameterizes  
different null rays. 
For the limit of small $r$ while keeping $t$ finite, $y$ goes to $\infty$. 
Then, in this limit, we find 
\begin{equation}
 {r\over r_{0\pm}}\approx y^{-3}={1\over 1-at/r}\approx {r\over
 a\xi}\, .
\end{equation}
Here we have defined $\xi=-t|_{r=0}$. 
From this relation, we can read that $r_{0\pm}=a\xi$. 
We consider a pair of an outgoing radial null ray and an incoming 
one such that the latter is the reflection of the former at the 
center. Then both rays have the same value of $\xi$, and hence 
we conclude that  
\begin{equation}
 r_{0+}=r_{0-}.
\label{ss1}
\end{equation} 

For a range of values of $\lambda$ the singularity forming in collapse
is naked. In this case,
the outgoing null ray emanating from the point at which the singularity 
first occurs, i.e., $r=0, t=0$, forms the Cauchy horizon. 
This null ray is given by $y=\alpha_4^-$ where we assume $\alpha_4^-$ 
is the largest real root among $\alpha_i^-$. 
Similarly, the incoming null ray terminating at the first singular point is 
also given by  $y=\alpha_4^+$ where, in the same way, $\alpha_4^+$ is the 
largest real root among $\alpha_i^+$.
Since we are considering the null rays close to the above limiting one,   
we expand the solution around $y=\alpha_4^{\pm}$. 
Then we find 
\begin{eqnarray}
 {r\over a \xi} & \approx & (y-\alpha_4^{\pm})^{-3A_4^{\pm}} 
          \prod_{i=1}^3
          (\alpha_4^{\pm}-\alpha_i^{\pm})^{-3A_i^{\pm}}\cr
      & \approx & C_{\pm} (t_{\pm}(r)-t)^{-3A_4^{\pm}}, 
\label{ss2}
\end{eqnarray}
where $t_{\pm}(r)=(r/a)[1-(\alpha_4^{\pm})^3]$ and 
\begin{eqnarray*}
  C_{\pm}=\left[{a\over 3r (\alpha_4^{\pm})^2} \right]^{-3A_4}
          \prod_{i=1}^3
          (\alpha_4^{\pm}-\alpha_i^{\pm})^{-3A_i^{\pm}}.
\end{eqnarray*}
As we have mentioned in the Introduction, we are considering a map between 
the sending time and the receiving time measured by an 
observer on a comoving shell. 
We denote the former time as $t_1$ and the latter time as $t_2$. 
Then, from (\ref{ss1}) and (\ref{ss2}), we obtain 
\begin{equation}
 (t_-(r_{sh}) -t_2)=\left[{C_{-}\over
     C_{+}}\right]_{ r=r_{sh}}^{1/(3A_4^-)}
   \hspace{-3mm}(t_+(r_{sh}) -t_1)^{A_4^+/A_4^-},  
\end{equation}
where $r_{sh}$ is the coordinate radius of the comoving shell 
on which the observer resides. 
Note that the result does not change even if we choose a very 
small value of $r_{sh}$. 
By a simple computation, we can show that $A_4^+/A_4^-$ here 
is the quantity denoted by $\gamma$ in Ref.\cite{SiVaWi}. 
Hence, we find that the basic property of the map around the 
singularity is maintained in this new calculation, which 
does not use any information about the geometry away from the 
first singular point.

\section{$\rho_2$-model}
In the preceding section, we considered the self-similar 
model in which the second derivative of the density profile at the 
center vanishes, and the third derivative is in a certain negative range. 
In this section, we would like to consider more general
cases in which the second derivative does not vanish.

What we would like to do here is 
to deduce an analogous map for the model with the $C^{\infty}$
initial density profile considered by Harada et al.\cite{Har1,Har2}
\begin{equation}
\rho{(R)} = \rho_{0} + \rho_{2} R^{2} + \cdots .
\end{equation}
We write the solution of Eq.(\ref{basic}) as
\begin{equation}
 R^3={9\over 4} F(r)(t-t_0(r))^2, 
\end{equation}
where $t_0(r)$ is an arbitrary function. 
By using the remaining gauge degrees of freedom 
we set 
\begin{equation}
 t_0(r)=r.  
\end{equation}
Then, the density profile near the center at some epoch $t=t_{in}<0$ 
before hitting the singularity, which first appears at $t=0, r=0$, 
is given by 
\begin{equation}
 \rho\approx {4\over 3 t_{in}^2}\left(1+{2\over t_{in}}
                 \left[r+{F\over F'}\right]\right)\, . 
\label{rhoapp}
\end{equation}
For small $r$, we assume $F(r)=\alpha r^\mu+O(r^\mu)$. 
Then $R^3\propto r^\mu$ for small $r$ at $t=t_{in}$. 
Requiring that the second term in the round brackets of Eq.(\ref{rhoapp}), 
$r+F/F'$, is proportional to $R^2$, we find that $\mu$ should be chosen 
as $3/2$. Then the condition $\rho_2<0$ is automatic, and the
singularity at the 
center turns out to be at least locally naked. 

Now let us consider a model truncated at the leading order for 
the expansion of $F(r)$, i.e., we suppose $F(r)=\alpha r^{3/2}$. 
Then the equation which determines the 
incoming and outgoing null geodesics is given by 
\begin{equation}
 {dt\over dr}
    =\mp {7\over 6}r^{1/6}{1-{3\over 7} t/r
           \over (1-t/r)^{1/3}}\, ,  
\label{geodesic1}
\end{equation}
where, for convenience, we have performed the following 
rescaling of variables,
\begin{eqnarray*}  
&& t\to \left(4\over 9\alpha \right)^{2} t, \cr
&& r\to \left(4\over 9\alpha \right)^{2} r. 
\end{eqnarray*}

For small $r$, the right hand side of the above equation goes to 
zero faster than $r^{-1}$. 
Hence the integration constant dominates the solution 
for small $r$ unless it vanishes. 
Thus, regarding the right hand side as small, 
we consider an expansion of the solution in the form 
\begin{equation}
 -t(r)=\xi-t_1(r) - t_2(r)-\cdots, 
\label{ansatz}
\end{equation}
where $\xi=-t|_{r=0}$ as before. 

To keep the notation simple, we introduce the function
\begin{equation}
{\cal F}(s)={1-{(3s/7)}\over (1-s)^{1/3}}\, .  
\end{equation}
Then, substituting the above ansatz (\ref{ansatz}) 
into Eq.(\ref{geodesic1}), we get the equations 
which determine $t(r)$ order by order:  
\begin{eqnarray}
{dt_1\over dr} & = & \mp {7\over 6}r^{1/6}{\cal F}(-\xi/r), \cr
{dt_2\over dr} & = & \mp {7\over 6}r^{1/6}
     {\cal F}'(-\xi/r){t_1\over r}\, , \cr
{dt_3\over dr} & = & \mp {7\over 6}r^{1/6}
     \left[{\cal F}'(-\xi/r){t_2\over r} 
          +{1\over 2}{\cal F}''(-\xi/r){t_1^2\over r^2}  \right]\, . 
\end{eqnarray}

Recall that 
\begin{equation}
{7\over 6} r^{1/6}{\cal F}(-\xi/r)={dR(-\xi,r)\over dr}\, , 
\end{equation}
which relation is not at all accidental but 
follows from the starting expression for the 
geodesic equation $dt/dr=\mp R'$.  
Using this relation, we can calculate $t_1(r)$ as 
\begin{equation}
 t_1(r)=\mp R(-\xi,r)=\mp r^{7/6}(1+\xi/r)^{2/3}. 
\end{equation}

The next order term is evaluated as 
\begin{eqnarray}
t_2 & = & {7\over 6}\int_0^r dr'\, 
        r'{}^{1/3}(1+\xi/r')^{2/3}{\cal F}'(-\xi/r')\cr
 & = & {7\over 6}\xi^{8/6} 
       \int_0^{r/\xi} d\mu\, \mu{}^{1/3}(1+1/\mu)^{2/3}{\cal F}'(-1/\mu)\, .
\end{eqnarray}

Expanding the integrand for large $r/\xi$, 
the asymptotic form of the integration  
is evaluated as 
\begin{eqnarray}
&& \int_0^{r/\xi} d\mu\, \mu{}^{1/3}(1+1/\mu)^{2/3}{\cal F}'(-1/\mu)\cr
  &&\qquad
 =C_2-{(r/\xi)^{4/3}\over 14}\left(1+{28\over 3}(\xi/r)%+\gamma_{2,2}(\xi/r)^2
          +\cdots \right)\, . 
\end{eqnarray}
Here $C_2$ is the integration constant, which cannot be determined 
by integrating the expression expanded for large $r/\xi$. 
%We do not show the explicit values of coefficients $\gamma_{i,j}$ 
%because they are not important here.  
Except for the term containing $C_2$,  
all other terms are completely determined 
by the asymptotic expansion of the integrand. 
One may notice that we have not introduced 
the corresponding constant for $t_1(r)$. In the case of $t_1(r)$ 
we can see that this constant vanishes by looking at the 
explicit expression which is written in terms 
of $R$. It will be worth stressing that this cancellation of the 
integration constant is not due to the assumed form of $R$. 

Also for $t_3(r)$, we can do a similar calculation as 
\begin{eqnarray}
 t_3 &=&\mp {7\over 6}\xi^{9/6} 
    \int_0^{r/\xi} d\nu\, \nu{}^{1/6} 
      \Biggl\{{7\over 6\nu}{\cal F}'(-1/\nu)
   \cr &&\quad\times 
      \int_0^\nu
         d\mu \mu^{1/3}(1+1/\mu)^{2/3}{\cal F}'(-1/\mu)\cr
&&\hspace{2cm}
   +{\nu^{1/3}\over 2} {\cal F}''(-1/\nu)(1+1/\nu)^{4/3}\Biggr\}\cr
& = & \mp {7\over 6}C_3 \xi^{9/6}
      \pm{7\over 9} C_2 \xi^{8/6} r^{1/6}(1+\cdots)\cr 
&&\qquad      \mp {11 r^{9/6}\over 162}
       \left(1 %+\gamma_{3,1}(\xi/r)+\gamma_{3,2}(\xi/r)^2
          +\cdots \right).
\end{eqnarray}
Besides the terms containing the integration constants $C_k$, 
terms in $t_k$ have the structure of $\xi^m r^{{k\over 6}-m+1} 
( m\ge 0 )$. 

The region of validity of the above expansion will be determined 
by comparing the first and second term. 
Requiring that $|t_1|/\xi \ll 1$, we find that the condition for 
the validity of the above expansion is satisfied for $r\ll \xi^{6/7}$. 

For sufficiently small $\xi$,  
within this region of validity, there is an overlapping region at which 
the condition $\xi/r\ll 1$ also holds. 
When this condition $\xi/r\ll 1$ is satisfied, 
we can expand the geodesic equation as 
\begin{equation}
 {dt\over dr}=\mp {7\over 6} r^{1/6}\left(
    1-{2t\over 21r}+{5t^2\over 63 r^2}+\cdots
    \right).   
\end{equation}
Then, it can be integrated iteratively as 
\begin{eqnarray}
 t -D_{\mp}
&=&r^{7/6}\cr
&&+\left(\pm{2\over 3}D_{\mp}r^{1/6}-{1\over 12}r^{8/6}\right)\cr
&&+\left(\pm{1\over 9}D_{\mp}^2 r^{-5/6}
    +{7\over 9}D_{\mp}r^{2/6}\mp {11\over 162}r^{9/6}\right)\hspace{-5mm}\cr
&&+\cdots,
\end{eqnarray}
where $D_{\mp}$ is the integration constant. 
Terms in the $(\ell+1)$-th line in the right hand side 
have the structure of $D_{\mp}^{\ell-n} r^{{7(n+1)\over 6}-\ell}\, 
(\ell\ge n\ge 0)$. Comparing them with the terms in $t_k$, we find 
the correspondence with $\ell=m+k-1,\, n=k$. Corresponding terms in 
two different expressions have the same coefficient. 
In other words, the first term in each line of the right hand side 
basically corresponds to $t_1(r)$, the second term to $t_2(r)$ and so on. 
The relation between $D_{\mp}$ and 
$\xi$ is found by comparing the coefficient of the $r$-independent term. 
Then we have 
\begin{eqnarray}
 D_{\mp}+O({D_{\mp}^2})& = & -\xi+{7\over 6}C_2 \xi^{8/6}
         \mp{7\over 6}C_3 \xi^{9/6}\cr
    &&\qquad +{7\over 6}C_4 \xi^{10/6}+O(\xi^{11/6}). 
\end{eqnarray}
Eliminating $\xi$ from this equation, we obtain the relation 
between $D_{+}$ and $D_{-}$, 
\begin{equation}
 D_{-}=D_{+}+{7\over 3} C_3(-D_{+})^{3/2}+O(D_{+}^{11/6}). 
\end{equation}
This mapping formula has the same structure of singularity as 
that obtained for the map between incoming and outgoing 
null coordinates by Harada, Iguchi and Nakao numerically.

\section{interpretation}
The map that we have obtained in this paper is the 
map between the ingoing and outgoing intersection points 
of null rays measured by the proper time of the comoving 
observer. This mapping is not exactly the same as the 
mapping of the incoming and outgoing null 
coordinates naturally defined for the observers 
at null infinities. However, as we mentioned in the Introduction, 
the singular behavior of these two mappings is expected to be 
common because there is no singular behavior in the null rays 
connecting the shell at a fixed $r$ and the shell at infinity. 

Then, in order to investigate the structure of the singular mapping, 
is it possible to set the location of shell $r_{sh}$ arbitrarily small? 
The answer is basically yes. 
In the above derivation, the only constraint on $r_{sh}$ is 
$r_{sh}\gg \xi$, but this constraint does not prevent us from choosing 
$r_{sh}$ arbitrary small.  
Since we are interested in the limiting behavior when 
the time of reflection $t=-\xi$ comes very close to the singularity
$t=0$, we can assume that $\xi$ is arbitrary small. 
In the preceding section, we considered a matching of two different 
expressions for the null geodesics. 
One is valid for small $r$ and the other for large $r$. 
At this moment, one may wonder why it was 
necessary to consider this matching although we are allowed 
to choose $r_{sh}$ arbitrarily small.
The reason is as follows. 
In the $\xi\to 0$ limit, 
the region of validity of the solution for small $r$ shrinks 
to zero. Hence, the solution for small $r$ cannot  
describe the limiting behavior for any value of $r_{sh}$.  

Anyway, the basic feature of the map is derived for 
arbitrarily small $r_{sh}$. This fact proves that it is determined 
by the geometry just around the central singularity as expected. 
This is unlike the case of the black hole for which the map is not 
determined by the local geometry near the center. 
Instead it is determined by the relationship between the 
regular interior coordinates and the external coordinates, 
near the event horizon.  

Here we have considered a specific model for the initial 
density profile. However, we can expect that the basic 
structure of the map will be independent of the details 
of the model because it is determined 
%the cancellation of $C_1$ occurs 
%independent of the initial density profile
by the geometry just near the point at which the singularity 
is first formed. 
In this small region, we would be able to neglect  
the higher order terms in the expansion of the density 
profile, although 
further consideration is necessary to give a rigorous 
proof for the statement that 
the structure of singularity in the mapping formula is 
totally determined by the leading term in the expansion 
of the initial density profile for general 
spherical dust collapse.

In this paper we have discussed models of 
spherically symmetric dust collapse. However, 
it would be of interest to generalize 
the prescription described in this paper to 
other matter models and to non-spherical collapse. 

\vspace{5mm}
\centerline{\bf Acknowledgements}
It is a pleasure to thank T. Harada H. Iguchi, A. Ishibashi,
H. Kodama and K. Nakao for useful 
discussions.
T.T.  acknowledges support from Monbusho Grant-in-Aid No. 1270154.
TPS acknowledges the partial support of the 
Funda\c{c}\~ao para a Ci\^encia e a Tecnologia (FCT),
Portugal under contract number SAPIENS/32694/99.

%\newpage

\end{document}